\documentclass{article}
\usepackage{spconf,amsmath,graphicx,hyperref}
\usepackage{booktabs, graphicx, multirow, enumitem, tabularx}
\usepackage[table]{xcolor}
\hypersetup{hidelinks}

\definecolor{hlblue}{RGB}{222,235,247}

\title{Hear Every Sound Object: \\Rescuing Overlooked Sources in Auditory Scene Analysis}
\name{Younghoo Kwon$^{1}$ \qquad Yongjoon Lee$^{1}$ \qquad Jung-Woo Choi$^{2 *}$\thanks{$^*$Corresponding author.}}
\address{$^{1}$School of Electrical Engineering, KAIST, Daejeon, South Korea \\
$^{2}$Department of AI Systems, KAIST, Daejeon, South Korea}
\begin{document}
\ninept
\maketitle
\begin{abstract}
Auditory scene analysis (ASA) models extract information about individual sound sources from a complex auditory mixture. Recent object-based ASA models process the mixture through a sequence of modules and split it into source-wise features (i.e., object features) only at the last layer. However, substantial disparities in performance across sources limit overall performance. Our analysis using centered kernel alignment (CKA) reveals that CKA scores correlate strongly with per-source performance, with their disparities widening from intermediate to deeper layers. These findings motivate the early split and exchange of object features to address source-wise performance disparities before they widen in deeper layers. To this end, we propose PARASITE, a plug-in refinement architecture that splits object features at earlier layers and progressively refines them through sequential layers. Each layer integrates object features separated at different depths using Object-wise Gated Fusion (OGF) and then refines them using Windowed Cross-Object Attention (WCOA). We introduce the DeepASA++ model, which combines PARASITE with an ASA backbone and delivers state-of-the-art results on both ASA2 and STARSS23. DeepASA++ achieves a SELD score of 0.240 on STARSS23 and surpasses prior multi-stage inference architectures, including chain-of-inference.
\end{abstract}

%
\section{Introduction}
\label{sec:intro}

Auditory scene analysis (ASA)~\cite{lee2025deepasaobjectorientedmultipurposenetwork} aims to characterize every sound source in a complex acoustic scene. The complex acoustic scene contains multiple simultaneously active sources, and each source is described by several properties: class, direction of arrival (DOA), frame-wise activity, and waveform. Prior studies have addressed these properties only in part. Sound event detection (SED) estimates the classes active in each time frame~\cite{khandelwal2024sound, schmid2024improving, yin2025exploring}, and sound event localization and detection (SELD) additionally estimates the DOA of each event from multichannel recordings~\cite{adavanne2018sound, dong2025experimental}. Source separation recovers the individual waveforms from the mixture~\cite{araki202530+, lee2025deft}. Although they target different properties, these models share one design principle: the network develops a single representation of the input mixture, which we refer to as the \emph{scene feature}, and task-specific decoders read the desired property directly from it, without ever forming an explicit representation of each source.

\begin{figure}[t]
    \centering
    \includegraphics[width=0.8\linewidth]{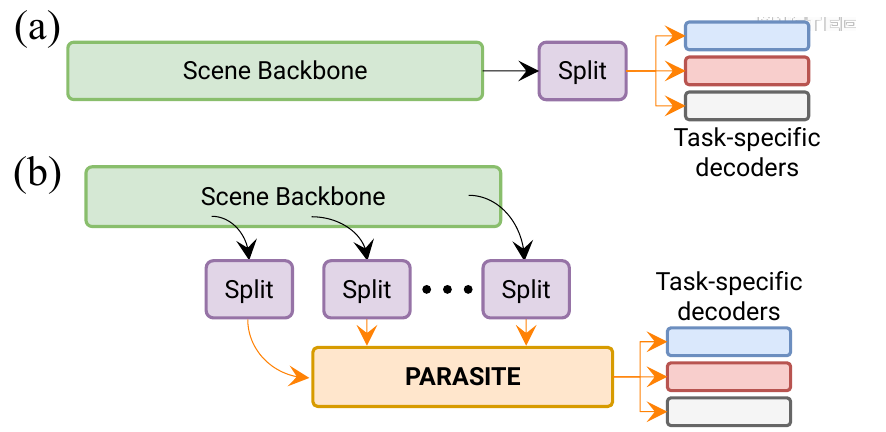}
    \caption{Comparison of split strategies. (a) Last-split strategy, as in DeepASA. (b) Proposed architecture with PARASITE.}
    \label{fig:concepts}
    \vspace{-0.4cm}
\end{figure}

DeepASA~\cite{lee2025deepasaobjectorientedmultipurposenetwork} unifies these tasks by splitting the scene feature into the feature that represents each individual source (i.e., object feature). As illustrated in Fig.~\ref{fig:concepts}(a), it consists of three parts, the first of which is the scene backbone. The scene backbone is a series of scene analysis blocks 
(DeFT-Mamba blocks~\cite{lee2025deft}) that develop the scene feature projected from the multichannel mixture into a form from which the sources can be readily separated. A split module then decomposes the scene feature into object features. Finally, task-specific decoders map each object feature to the corresponding task output. DeepASA thus follows a \textit{last-split} strategy: only the scene feature at the last layer of the backbone is split into object features.

Under the last-split strategy, performance varies widely across sources within a single mixture, with certain sources estimated far less accurately than others.  We hypothesize that this variation comes from how \textit{dominant} each source is in the scene feature, which jointly represents all sources. To quantify each source's dominance in the scene feature, we measure layer-wise centered kernel alignment (CKA)~\cite{kornblith2019similarityneuralnetworkrepresentations} between the scene feature and a reference representation (e.g., Mel-spectrogram or SSL representation) for each source. As detailed in the next section, the ASA performances of sources show a strong correlation with their CKA values. Moreover, the CKA gap between sources is smallest at the middle layers and widens dramatically at the final layer. The large performance gap between sources observed in the last-split strategy motivates investigating whether intermediate features can improve less dominant sources.



\begin{figure*}[t]
    \centering
    \vspace{-0.2cm}
    \includegraphics[width=0.95\linewidth]{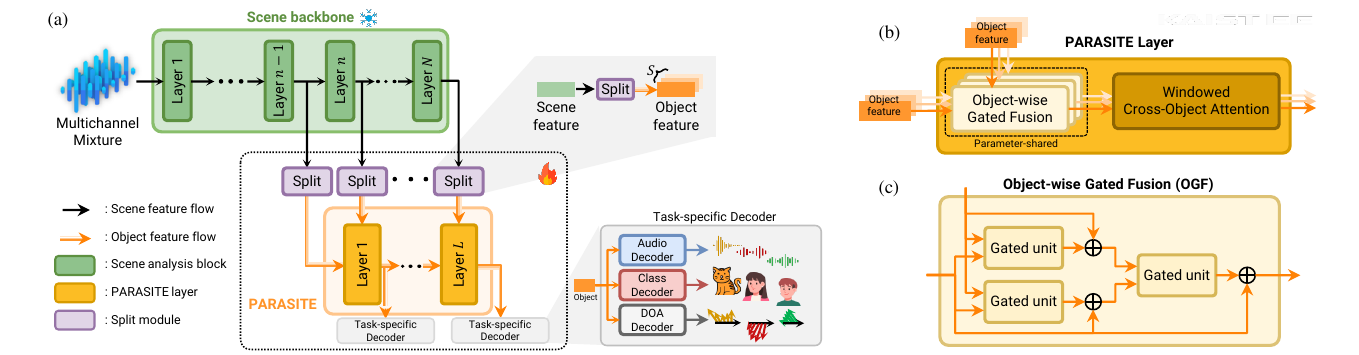}
    \caption{An overall framework of DeepASA++. (a) PARASITE develops the object features alongside the scene backbone. (b) A PARASITE layer: Object-wise Gated Fusion (OGF) followed by Windowed Cross-Object Attention (WCOA). (c) OGF built from three gated units.}
    \label{fig:architecture}
\end{figure*}

To resolve the problem, we attach \textbf{PARASITE}, \textbf{P}lug-in \textbf{A}uxiliary \textbf{R}efinement \textbf{A}rchitecture for \textbf{S}patial \textbf{I}nference and \textbf{T}emporal \textbf{E}vent detection, to the scene backbone, as illustrated in Fig.~\ref{fig:concepts}(b). PARASITE is designed to preserve information about less dominant sources by using object features split from middle-layer scene features, where the CKA gap between sources is small. PARASITE sequentially refines object features using new object features split from subsequent scene features, up to the final layer. Specifically, PARASITE is a series of blocks built from an object-wise module and a cross-object module, as shown in Fig.~\ref{fig:architecture}(b). Object-wise Gated Fusion (OGF) processes each object individually, fusing the object feature carried from the previous block with the object feature split from the current scene feature. Since OGF keeps the object features separate, a module that lets the object features interact is required. The modules proposed for this purpose~\cite{shin2024separate, lee2024boosting, shin2026asymmetric} exchange information only within the same time frame. We therefore introduce Windowed Cross-Object Attention (WCOA) that allows object features to exchange information over a temporal window rather than a single time--frequency bin. 


Combining these components, we present \textbf{DeepASA++}, an ASA model with PARASITE integrated into DeepASA-Loco, which uses TF-Locoformer blocks~\cite{saijo2024tflocoformertransformerlocalmodeling} as the scene backbone. The demos and code are available at the following link\footnote{\url{https://sites.google.com/view/deepasapp/}}. The contributions of this work are summarized as follows:
\begin{itemize} [leftmargin=1.2em,itemsep=1pt,topsep=2pt,parsep=0pt]
    \item We find that (1) CKA-based source rankings are associated with per-source performance, and (2) the CKA gap between sources is smallest at the middle layers and widens at the last layer. 

    \item To use middle-layer features with small CKA gaps, we split object features at these layers and pass them through an additional branch (PARASITE) with the proposed modules.
    
    \item Our proposed system reaches SOTA performance on ASA2 and STARSS23, achieving 14.24 dB SI-SDRi and a SELD score of 0.147 on ASA2 and 0.240 on STARSS23.
\end{itemize}
\vspace{-0.1cm}

\section{Source dominance analysis}
\label{sec:findings}

\noindent \textbf{Methods.} CKA~\cite{kornblith2019similarityneuralnetworkrepresentations} is a statistical kernel alignment method that measures the similarity between two features. To investigate the performance gap between the sources, we measure the CKA between the scene features at each layer of DeepASA-Loco and a reference feature extracted from the target of each source. We use two reference features: the Mel-spectrogram and the SPEAR embedding~\cite{yang2026spearunifiedsslframework}, ensuring both the acoustic and semantic aspects of a source are covered. The analysis proceeds in two steps. First, we rank sources within each mixture to compare source dominance. We average each source's CKA values across all layers and order sources by these values, referring to the order as \textbf{dominance rank} (rank~1: the most dominant source). Second, we group sources across mixtures by dominance rank and compute a CKA value for each group at each layer. We carry out the analysis on mixtures from the ASA2 training split, each containing five sources (about 19K mixtures). To assess robustness, we draw 50 random subsets of 500 mixtures each and report the mean and 95\% confidence interval across these subsets.

\begin{figure}[!ht]
    \centering
    \begin{minipage}[t]{0.49\columnwidth}
        \centering
        \includegraphics[width=\linewidth]{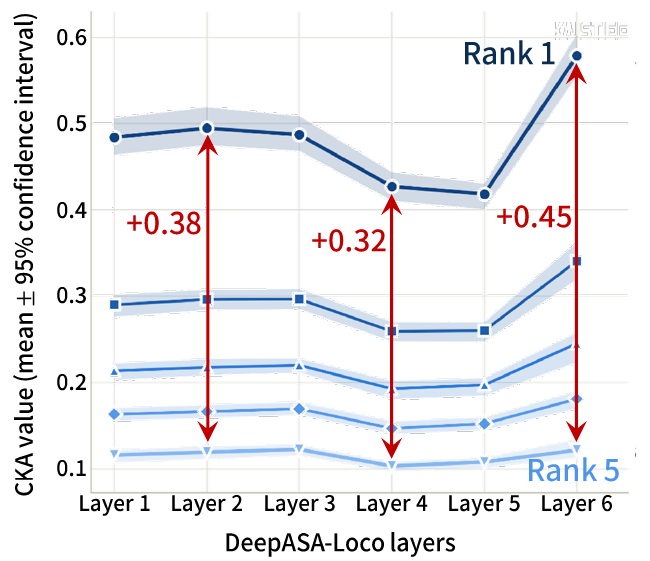}
        \vspace{-1mm}
        \centerline{(a) Mel-spectrogram}
    \end{minipage}
    \hfill
    \begin{minipage}[t]{0.49\columnwidth}
        \centering
        \includegraphics[width=\linewidth]{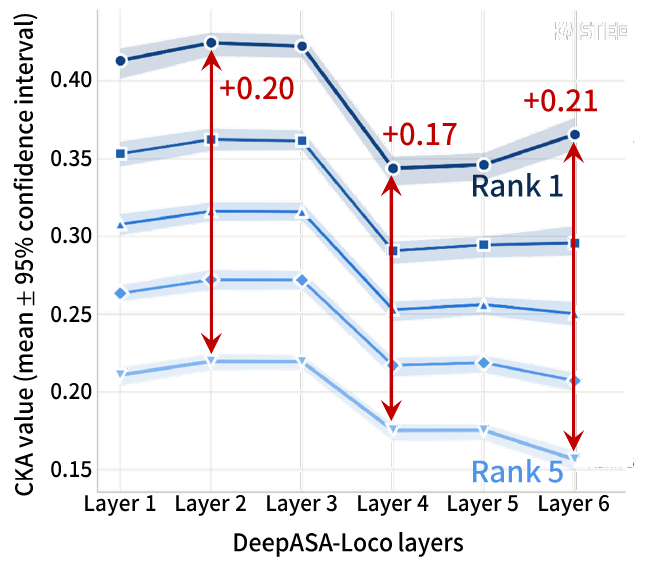}
        \vspace{-1mm}
        \centerline{(b) SPEAR}
    \end{minipage}\caption{Layer-wise CKA between the scene feature of DeepASA-Loco and each embedding of the isolated source.}
    \label{fig:findings}
\end{figure}

\noindent \textbf{The widening CKA gap toward the final layer suggests that scene features may favor dominant sources, potentially leaving less dominant sources insufficiently represented.} Fig.~\ref{fig:findings} shows the result. From layer~4 to the last layer, the CKA gap between the highest and lowest dominance ranks grows from 0.32 to 0.45 with the Mel-spectrogram, and the SPEAR embedding follows the same trend, from 0.17 to 0.21. A widening gap in CKA between the most and least dominant sources means that the dominant source strengthens its hold on the scene feature while the least dominant source is progressively suppressed, and this effect is most pronounced at the final scene feature. Under the last-split strategy, that final representation is the only one the split module ever sees, so the least dominant object feature is decoded from the representation in which that object is least present. Table~\ref{tab:rank} shows the associated performance gap: with the Mel-spectrogram, the rank~5 sources fall 10.60 dB behind the rank~1 sources in SI-SDR, 21.32$^\circ$ in localization error, and 27.4\%p in classification accuracy. This motivates forming the object feature from Layer~4 onward, where the gap between the ranks is smallest, and carrying them forward so that the information about the less dominant sources survives to the output.

\begin{table}[t]
\centering
\caption{DeepASA-Loco performance by dominance rank, with Mel-spectrogram and SPEAR embedding as the reference feature. Source gap represents the performance gap between ranks 1 and 5.}
\label{tab:rank}
\setlength{\tabcolsep}{4pt}
\renewcommand{\arraystretch}{0.95}
\scriptsize
\begin{tabular}{clcccccc}
\toprule
& & \multicolumn{5}{c}{Dominance rank} & \\
\cmidrule(lr){3-7}
Repre. & Metric & \begin{tabular}[c]{@{}c@{}}1\\(most)\end{tabular} & 2 & 3 & 4
& \begin{tabular}[c]{@{}c@{}}5 \\ (least)\end{tabular}
& \begin{tabular}[c]{@{}c@{}}Source gap\end{tabular} \\
\midrule
\multirow{3}{*}{Mel} & SI-SDR (dB)\,$\uparrow$   & 9.57 & 6.60 & 4.62 & 2.50 & $-$1.03 & $+$10.60 \\
& LE ($^\circ$)\,$\downarrow$ & 8.92 & 13.72 & 17.32 & 22.22 & 30.24 & $-$21.32 \\
& Acc. (\%)\,$\uparrow$ & 97.9 & 92.9 & 89.1 & 83.3 & 70.5 & $+$27.4 \\
\midrule
\multirow{3}{*}{SPEAR} & SI-SDR (dB)\,$\uparrow$   & 8.12 & 5.24 & 3.74 & 2.44 & 2.72 & $+$5.40 \\
& LE ($^\circ$)\,$\downarrow$ & 11.15 & 16.52 & 20.15 & 22.95 & 21.63 & $-$10.48 \\
& Acc. (\%)\,$\uparrow$ & 95.7 & 89.4 & 85.2 & 81.0 & 82.3 & $+$13.4 \\
\bottomrule
\end{tabular}
\end{table}

\section{Methodology}
\label{sec:pagestyle}
\vspace{-0.1cm}
As shown in Fig.~\ref{fig:architecture}(a), the proposed DeepASA++ consists of the scene backbone with $N$ layers, the split modules, and PARASITE with $L$ layers followed by task-specific decoders. The scene backbone, trained with the last-split strategy, processes scene features. The scene feature has a shape of $D$, $T$, $F$ where $D$, $T$, and $F$ denote the channel, time, and frequency dimensions. The split module expands the channel dimension by a factor of $S$ using a $3\times3$ convolution layer, where $S$ is the maximum number of objects, while leaving the time and frequency dimensions unchanged. Splitting the expanded feature along the channel dimension yields $S$ object features, each having the same shape as the scene feature. This split module is attached to each of the last $L+1$ scene analysis blocks, so that every layer of PARASITE is supplied with its own set of object features. PARASITE processes these object features, with each layer sequentially fusing the previous layer's object features with the object features split from the current scene feature. The first PARASITE layer instead fuses the object features split from two adjacent scene features. As shown in Fig.~\ref{fig:architecture}(b), a layer of PARASITE consists of two complementary modules: Object-wise Gated Fusion (OGF), which processes each object on its own, and Windowed Cross-Object Attention (WCOA), which lets the objects interact by looking up adjacent time frames. 
 
OGF fuses the object features representing the same object: the one from the previous layer and the one split from the current scene feature. OGF consists of three gated units shown in Fig.~\ref{fig:architecture}(c). At first, two gated units take the same pair of inputs but use different skip connections, so that each unit produces a refined feature from each input. The last gated unit then fuses these two refined features into the OGF output. The final output has a skip connection to the object feature from the previous layer, so that the OGF output remains anchored to the representation carried over from the previous layer. A gated unit concatenates the two object features along the channel axis and applies a point-wise convolution, layer normalization, a Swish activation, and a $7\times7$ depth-wise convolution without any dimension change. The gated unit then splits the output in half along the channel axis and passes one half through a sigmoid to gate the other by element-wise multiplication. The information is not exchanged between object features because both inputs describe the same object.
WCOA provides that exchange. WCOA performs multi-head self-attention along the object axis, taking $S$ object features from OGF as its input. WCOA treats every time--frequency bin $(t,f)$ of every object as a token, where $t$ and $f$ index the time and frequency dimensions. For each query at bin $(t,f)$ of one object, the keys are collected from all $S$ objects over a temporal window of $w$ bins centered on $(t,f)$, yielding $S\times w$ logits. After the softmax, the probabilities belonging to the same object are summed, so that each object receives a single attention weight that is applied to its value at the query's own bin $(t,f)$. The temporal window, therefore, serves as evidence for weighting the objects so that a query can account for interference spread over neighboring frames. 

\section{Experimental setups}
\label{sec:experiment}
\vspace{-0.1cm}
\noindent\textbf{Datasets.}
We used ASA2~\cite{lee2025deepasaobjectorientedmultipurposenetwork} for both training and evaluation. The mixtures in the dataset contain up to $S=5$ sources from 13 classes. We also used STARSS23~\cite{shimada2023starss23} to evaluate SELD performance. For STARSS23, the model was trained on the augmented data of~\cite{lee2025deepasaobjectorientedmultipurposenetwork}.
\vspace{0.1cm}

\noindent\textbf{Model configurations.} We trained DeepASA-Loco, which follows the DeepASA framework~\cite{lee2025deepasaobjectorientedmultipurposenetwork} but uses TF-Locoformer blocks~\cite{saijo2024tflocoformertransformerlocalmodeling} as its scene analysis blocks. 
DeepASA-Loco consists of $N=6$ TF-Locoformer blocks with a channel dimension of $D=64$, where the Conv-SwiGLU modules use a channel expansion factor of 2. We added PARASITE consisting of $L=3$ layers to use scene features with a small CKA gap. We set the number of objects to $S=5$ and the number of temporal window frames to $w=5$ for WCOA.
\vspace{0.1cm}

\noindent\textbf{Training configurations on ASA2.} Training was conducted in two stages. In the first stage, DeepASA-Loco was trained from scratch following the last-split strategy for 150 epochs. In the second stage, PARASITE was trained with every module of the first stage frozen. We decode every object feature of each PARASITE layer with individual decoders and compute the loss for each, following~\cite{shin2024separate, olsen2026knowing}, but use only the last-block decoder outputs at inference. We set the permutation for minimizing the loss at the last layer following~\cite{olsen2026knowing}. The linear warm-up was set to 4K steps, and the other configurations follow DeepASA~\cite{lee2025deepasaobjectorientedmultipurposenetwork}.



\vspace{0.1cm}
\noindent\textbf{Training configurations on STARSS23.} Starting from the model trained on ASA2, training was again conducted in two stages. In the first stage, DeepASA-Loco was fine-tuned on STARSS23. In the second stage, PARASITE was fine-tuned on top of this backbone, which was kept frozen. The fine-tuning configurations follow DeepASA~\cite{lee2025deepasaobjectorientedmultipurposenetwork}, with additional linear warm-up for 4K steps.
\vspace{0.1cm}


\noindent\textbf{Evaluation metrics.} Our model estimates the direct sound and the reverberant component of each source at every output channel, so we report Scale-Invariant Signal-to-Distortion Ratio improvement (SI-SDRi)~\cite{roux2018sdrhalfbakeddone} measured on the sum of the two estimates and direct SI-SDRi measured on the direct sound alone. For detection and localization, we used the SELD score~\cite{adavanne2018sound}. 

\section{Results}
\label{sec:results}
\vspace{-0.1cm}
\begin{table*}[t]
\centering
\caption{Results on ASA2. The results of the models with \textsuperscript{*} are taken from~\cite{lee2025deepasaobjectorientedmultipurposenetwork}. The other models are retrained. DeepASA and DeepASA-Loco utilize DeFT-Mamba~\cite{lee2025deft} and TF-Locoformer~\cite{saijo2024tflocoformertransformerlocalmodeling} blocks as 
scene analysis blocks, respectively.}
\label{tab:main}
\scriptsize
\setlength{\tabcolsep}{4pt}
\renewcommand{\arraystretch}{0.95}
\begin{tabular}{lccccccccc}
\toprule
& \multicolumn{2}{c}{Separation}
& \multicolumn{5}{c}{Detection and localization} \\ 
\cmidrule(lr){2-3}\cmidrule(lr){4-8} 
Model
& \begin{tabular}[c]{@{}c@{}}SI-SDRi (dB)\,$\uparrow$\end{tabular}
& \begin{tabular}[c]{@{}c@{}}direct SI-SDRi (dB)\,$\uparrow$\end{tabular}
& \begin{tabular}[c]{@{}c@{}}SELD score\,$\downarrow$\end{tabular}
& \begin{tabular}[c]{@{}c@{}}ER (\%)\,$\downarrow$\end{tabular}
& \begin{tabular}[c]{@{}c@{}}F1 (\%)\,$\uparrow$\end{tabular}
& \begin{tabular}[c]{@{}c@{}}LE ($^\circ$)\,$\downarrow$\end{tabular}
& \begin{tabular}[c]{@{}c@{}}LR (\%)\,$\uparrow$\end{tabular} \\ 
\midrule
SpatialNet\textsuperscript{*}~\cite{quan2024spatialnet}          & 9.60   & --    & --    & --   & --   & --   & --   \\ 
DeFT-Mamba\textsuperscript{*}~\cite{lee2025deft}          & 10.40  & --    & --    & --   & --   & --   & --   \\ 
SELD-Mamba\textsuperscript{*}~\cite{mu2025seld}          & --    & --    & 0.396 & 43.5 & 42.7 & 25.5 & 56.7 \\ 
MFF-EINV2\textsuperscript{*}~\cite{mu2024mff}           & --    & --    & 0.381 & 42.1 & 43.2 & 25.8 & 60.7 \\ 
DeepASA~\cite{lee2025deepasaobjectorientedmultipurposenetwork} & 12.38 & 11.91 & 0.190 & 24.5 & 79.1 & 16.1 & 78.2 \\ 
\quad w/ PARASITE & 13.84 & 13.75 & 0.165 & 22.1 & 81.5 & 13.0 & 81.9 \\
\midrule
DeepASA-Loco & 13.88 & 13.78 & 0.176 & 22.8 & 78.9 & 15.0 & 81.8 \\ 
\quad w/ CoI & 13.64 & 14.06 & 0.183 & 24.8 & 76.2 & \textbf{10.5} & 81.5 \\
\rowcolor{hlblue} \quad w/ PARASITE (DeepASA++)
  & \textbf{14.24}
  & \textbf{14.64}
  & \textbf{0.147}
  & \textbf{19.2} & \textbf{83.3} & 12.5 & \textbf{84.0} \\ 
\bottomrule
\end{tabular} 
\end{table*}

Table~\ref{tab:main} reports the results on ASA2, where all tasks are evaluated at once. Our reproduced DeepASA, whose scene analysis blocks are DeFT-Mamba~\cite{lee2025deft}, outperforms models specialized in a single task. DeepASA exceeds DeFT-Mamba by 1.98\,dB in SI-SDRi, and its SELD score is 0.191 lower than MFF-EINV2~\cite{mu2024mff}. Replacing the scene analysis blocks with TF-Locoformer blocks~\cite{saijo2024tflocoformertransformerlocalmodeling} improves most metrics, resulting in DeepASA-Loco.
Chain-of-Inference (CoI) module~\cite{lee2025deepasaobjectorientedmultipurposenetwork}, an auxiliary module that refines object features following the last-split strategy, does not consistently improve performance.
Attaching PARASITE to DeepASA-Loco yields DeepASA++, which achieves 14.24 dB SI-SDRi and a SELD score of 0.147, the best in both separation and SELD. Attaching the same PARASITE to DeepASA improves performance, indicating that the gain does not depend on the backbone choice.


Table~\ref{tab:ablation} reports ablation studies. The proposed model, DeepASA++, is highlighted in the table. For a fair comparison, we doubled the channel expansion rate in the Conv-SwiGLU~\cite{saijo2024tflocoformertransformerlocalmodeling} from 2 to 4 for DeepASA-Loco (i.e., DeepASA-Loco (large)). The results show that increasing model size does not guarantee performance improvement. DeepASA-Loco (large) uses 1.69 times the parameters and 1.53 times the computation of DeepASA++ while remaining 0.45\,dB behind in SI-SDRi and 0.017 behind in SELD score. 


\vspace{0.1cm}
\noindent\textbf{Layer selection.} The `Splits' column in Table~\ref{tab:ablation} identifies the scene analysis blocks providing current scene features to successive PARASITE layers, excluding the initial split. For DeepASA-Loco, it denotes the layer used for final decoding. Motivated by their small CKA gaps between sources, we selected Layers~4 and~5 alongside the final layer for the proposed configuration. Restricting the current scene input to Layer~6 yields only a marginal improvement in SELD score over DeepASA-Loco (0.176~$\rightarrow$~0.172), highlighting the benefit of incorporating earlier object features. We further examined the layer-selection criterion by replacing Layers~4 and~5 with Layers~2 and~3, where the CKA gaps are larger, while retaining Layer~6. Although this alternative outperforms the configuration using only Layer~6 as the current scene input, it remains inferior to the proposed configuration. These comparisons suggest that layers with small CKA gaps are the most effective points for extracting object features among the tested configurations.
\vspace{0.1cm}



\noindent\textbf{Modules in PARASITE.} To show the effectiveness of the proposed modules in PARASITE, we replaced or removed each module. First, we replaced OGF with a single gated unit connected to the object feature of the previous layer by a skip connection. The variant degrades the SELD score, even compared with DeepASA-Loco (0.176~$\rightarrow$~0.180). This illustrates that the way to fuse object features is important, even when the features from early layers are used. Second, removing WCOA also degrades both metrics compared with DeepASA++. This indicates that exchanging information between object features is essential for refining the object features, since OGF processes each object feature independently.



\begin{table}[t]
\centering
\vspace{-0.4cm}
\caption{Ablation studies on ASA2 about layer selection and proposed modules in PARASITE.}
\label{tab:ablation}
\setlength{\tabcolsep}{4.5pt}
\renewcommand{\arraystretch}{0.5}
\scriptsize
\resizebox{\columnwidth}{!}{
\begin{tabular}{lccccc}
\toprule
Model & Splits & SI-SDRi (dB)\,$\uparrow$ & SELD\,$\downarrow$ & Param. (M) & GMAC/s \\
\midrule
DeepASA-Loco & 6 & 13.88 & 0.176 & 6.10 & 64.21 \\
DeepASA-Loco (large)  & 6 & 13.79 & 0.164 & 11.89 & 159.64 \\
\midrule
DeepASA++         & 6       & 14.08             & 0.172  & 6.40   & 77.79 \\
DeepASA++         & 2,~3,~6 & 14.15             & 0.158  & 7.02   & 104.49    \\
\rowcolor{hlblue} DeepASA++        & 4,~5,~6 & \textbf{14.24} & \textbf{0.147} & 7.02 & 104.49 \\
\quad OGF~$\rightarrow$~A gated unit   & 4,~5,~6 & 13.98 & 0.180 & 6.87 & 92.01 \\
\quad w/o WCOA  & 4,~5,~6 & 14.06 & 0.170 & 6.85 & 90.50 \\

\bottomrule
\end{tabular}}
\end{table}

\begin{figure}[t]
\centering
    \begin{minipage}[t]{0.49\columnwidth}
        \centering
        \includegraphics[width=\linewidth]{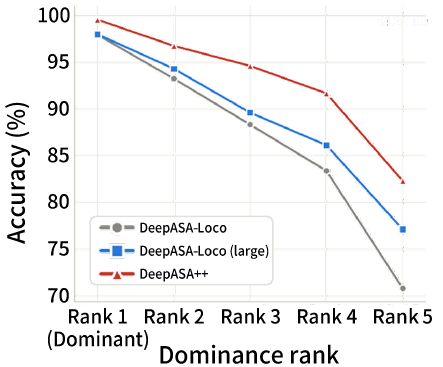}
        \vspace{-1mm}
        \centerline{(a) Classification accuracy}
    \end{minipage}
    \hfill
    \begin{minipage}[t]{0.49\columnwidth}
        \centering
        \includegraphics[width=\linewidth]{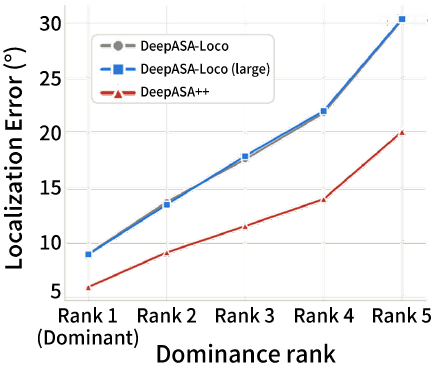}
        \vspace{-1mm}
        \centerline{(b) Localization error}
    \end{minipage}
    \caption{Per-source performance by dominance rank on ASA2.}
    \vspace{-0.1cm}
    \label{fig:performance_gap}
\end{figure}

Fig.~\ref{fig:performance_gap} presents classification accuracy and localization error by dominance rank. We compute ranks from DeepASA-Loco scene features using Mel-spectrograms as the reference representation. We use data from the ASA2 dataset, including exactly five sources. Across all models, performance degrades monotonically from rank~1 to rank~5. Enlarging DeepASA-Loco improves classification only for low-ranked sources and yields almost no gain in localization. In contrast, DeepASA++ clearly outperforms both models at every rank in both metrics. Notably, the improvement of DeepASA++ over DeepASA-Loco grows steadily from rank 1 to rank 5. This confirms that splitting object features at middle layers, where the CKA gap between ranks is small, benefits less dominant sources the most, as intended by PARASITE.

\begin{table}[t]
\vspace{-0.4cm}
\caption{SELD performance on STARSS23. For models with~\textsuperscript{*}, results are taken from~\cite{lee2025deepasaobjectorientedmultipurposenetwork}.}
\centering
\label{tab:seld}
\setlength{\tabcolsep}{3pt}
\scriptsize
\renewcommand{\arraystretch}{0.95}
\begin{tabular}{lccccc}
\toprule
Model & SELD\,$\downarrow$ & ER (\%)\,$\downarrow$ & F1 (\%)\,$\uparrow$ & LE ($^\circ$)\,$\downarrow$ & LR (\%)\,$\uparrow$ \\
\midrule
CST-former2\textsuperscript{*}~\cite{shul2025cst}             & 0.301 & 42.0 & 59.7 & 15.6 & 68.4 \\
EINV-2 (2nd rank)\textsuperscript{*}~\cite{xue2023attention}       & 0.301 & 42.0 & 57.5 & 15.8 & 72.7 \\
NERC-SLIP (1st rank)\textsuperscript{*}~\cite{Du_NERCSLIP_task3_report}    & 0.277 & 40.0 & 64.0 & 13.4 & 74.0 \\
\quad w/ Ensemble\textsuperscript{*}~\cite{Du_NERCSLIP_task3_report}               & 0.260 & 38.0 & 66.0 & 12.8 & \textbf{75.0} \\
DeepASA\textsuperscript{*}~\cite{lee2025deepasaobjectorientedmultipurposenetwork}                 & 0.259 & 34.4 & 62.6 & 10.2 & 73.8 \\
\quad w/ CoI\textsuperscript{*}~\cite{lee2025deepasaobjectorientedmultipurposenetwork}            & 0.253 & 33.7 & 63.1 & 9.8 & 74.6 \\
\midrule
DeepASA-Loco      & 0.257 & 35.4 & 65.1 & 10.0 & 73.1 \\
\rowcolor{hlblue} DeepASA++    & \textbf{0.240} & \textbf{33.5} & \textbf{67.5} & \textbf{8.4} & 74.5 \\
\bottomrule
\end{tabular}
\end{table}

Table~\ref{tab:seld} reports the SELD performance on STARSS23~\cite{shimada2023starss23}, where no separation target is available, and only the detection and localization objectives are used. DeepASA++ achieves the best SELD score of 0.240, 0.013 below the previous best~\cite{lee2025deepasaobjectorientedmultipurposenetwork}. It also outperforms the top-ranked DCASE submission with an ensemble~\cite{Du_NERCSLIP_task3_report} by 0.020. Attaching PARASITE to DeepASA-Loco yields a larger improvement in SELD score (0.257~$\rightarrow$~0.240) than attaching CoI to DeepASA (0.259~$\rightarrow$~0.253).
This shows that PARASITE is also more effective than CoI~\cite{lee2025deepasaobjectorientedmultipurposenetwork}, which follows the last-split strategy. The gain is large in localization: the localization error drops from 10.0$^\circ$ to 8.4$^\circ$, the lowest among the compared systems.

\section{Conclusion}
\vspace{-0.1cm}


To address the substantial performance disparities across sources in object-based ASA models, we proposed PARASITE and DeepASA++, which extract object features early and exchange information with features derived from the scene backbone. Motivated by the observation that source-wise CKA scores exhibit small disparities at intermediate layers but large disparities at the final layer, PARASITE extracts object features at earlier layers and integrates them with those extracted from the scene backbone at subsequent depths. By progressively refining object features through Object-wise Gated Fusion and Windowed Cross-Object Attention, DeepASA++ achieves state-of-the-art performance on the ASA2 and STARSS23 datasets. Consistent with our motivation, splitting object features at layers with a small CKA gap brings the largest gain, especially for less dominant sources.





\section{Acknowledgements}
{\footnotesize
This work was supported by the National Research Foundation of Korea (NRF) grant (No. RS-2024-00337945), the STEAM research grant (No. RS-2024-00464269) funded by the Ministry of Science and ICT of Korea government (MSIT), the BK21 FOUR program through the NRF grant funded by the Ministry of Education of Korea government (MOE), and Artificial intelligence industrial convergence cluster development project funded by the Ministry of Science and ICT (MSIT, Korea) \& Gwangju Metropolitan City.
}
{\footnotesize
\bibliographystyle{IEEEbib}
\bibliography{strings,refs}}

\end{document}